# A Complexity Agnostic Clustering Engine for Time Projection Chambers and its Implementation in FPGA

Jinyuan Wu, Michael Wang and Datao Gong

*Abstract*— A clustering functional block implemented in field-programable-gate-array (FPGA) for time projection chambers (TPC) operating with predictable time regardless the complexity of the event is described in this paper. The clustering functional block reorganizes input data and the hits data belonging to the same clusters are output together for further process in the later stages. The clustering operation consists of two phases, data filling phase and data outputting phase, and the later uses the same number of clock cycles as the data filling phase. The clustering block can accommodate events with arbitrary number of clusters and number of hits per cluster as long as the total number of hits is within a predesigned limit. The operation time is exactly twice of the data filling time with no residual $O(n^2)$ term. The clustering block has been implemented with operating frequency of 200 MHz in a low-cost FPGA evaluation module and test results confirm the expected performance.

*Index Terms*— Trigger Systems, FPGA Applications, Time Projection Chamber, Clustering

## I. Introduction

In modern high energy physics and nuclear physics experiments, time projection chambers (TPC), either gas-filled or liquid Argon based ones, are widely used. In TPCs, clustering is a common task which is to organize data belonging the same charged particle track together for further analysis. The clustering is done in software essentially with two layers nested-loops, consuming $O(n^2)$ time, where n is number of hits in an event.

For real-time trigger systems, clustering tasks must be performed in $O(n)$ time. In mainstream clustering schemes implemented in Field-Programable-Gate-Arrays (FPGA), multiple copies of hardware blocks are used to "unroll" a layer of the nested-loops. These direct unrolling approaches are usable when the complexity of the event is within certain limits, for example, when number of hits per event, number of clusters per event or number of hits per cluster are limited. When processing high complexity events, the direct unrolling schemes may either fail directly or use non-deterministic operating time with residual $O(n^2)$ terms.

The clustering operation is a type of doublet finding process, and therefore, it intrinsically should be possible to find an approach using indexing search to eliminate any computation requiring $O(n^2)$ time. Block random access memories (RAM) available in both CPU and FPGA computation environments are suitable platform for indexing search operations.

The clusters in TPC are usually in 2-dimensional channel-time space as shown in Fig. 1(a). Hits in the neighboring channels with either same or adjacent timing bins are considered belonging to the same cluster.

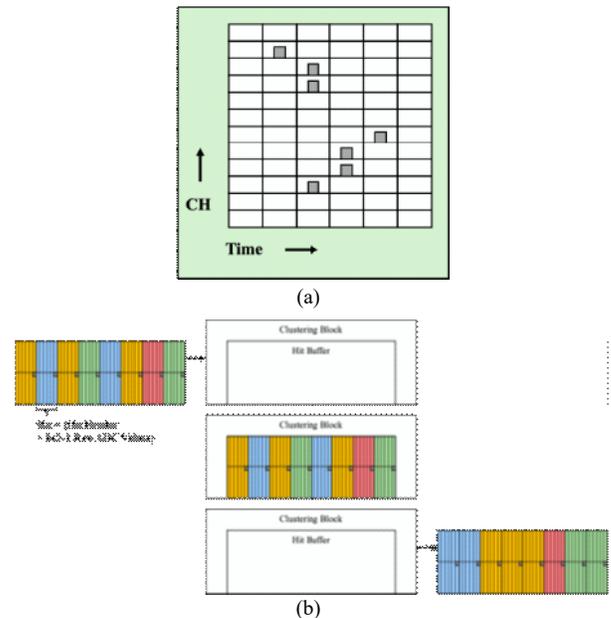

Fig. 1. Clustering process of the TPC data stream

In TPC data stream, a hit is usually a collection of ADC values representing a waveform when the cloud of electrons drifting to the collection pad or passing by the inducing wire. It would be desirable for the clustering block to accept and preserve the full data package during the clustering process. A typical clustering process for a TPC stream is shown in Fig. 1(b).

In the example above, a hit contains a hit header plus 15 (or more) ADC values, taking 8 (or more) clock cycles to feed into the clustering block. After all hits in an event is filled into the clustering block, they are output from it with essentially the same data structure as the input, except the hits are re-ordered so that all hits belonging a cluster are output together.

In this document, the principle of the clustering block is first discussed in Section II. FPGA implementation and test results from outputs of the FPGA are presented in Section III followed with conclusion in Section IV.

## II. PRINCIPLE OF THE CLUSTERING ENGINE

The block diagram of the clustering engine is shown in Fig. 2.

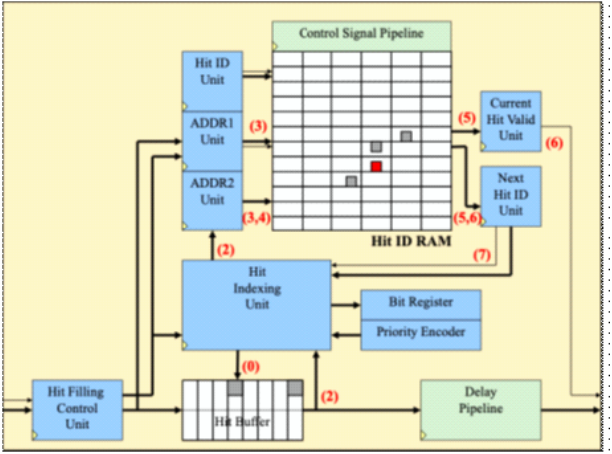

Fig. 2. The block diagram of the clustering engine

The center piece of the clustering block is a block memory called Hit ID RAM. The Hit ID RAM is organized as a two-dimensional array, where the X-axis represents the time bin and the Y-axis represents the channel.

The time bin size is chosen carefully: it must be small enough to ensure that no double hits occur within a single cell, yet large enough to guarantee that hits belonging to the same cluster remain connected.

A cluster forms a simple line in the Time–CH map. From the current hit to the next hit, the channel (CH) can change by +1 or −1, and the time bin can change by +1, 0, or −1. During the clustering process, the CH+1 hit is followed first. The CH−1 hit is processed only after all CH+ hits have been addressed. If hits are disconnected or branched, the additional hit will be treated as a separate cluster and therefore, no hits will be discarded.

The first phase of the clustering starts with data filling. The hit header and raw ADC data are fed into the Hit Buffer of the clustering engine and stored until output phase. The data format of a hit is shown in the left side of Fig. 3.

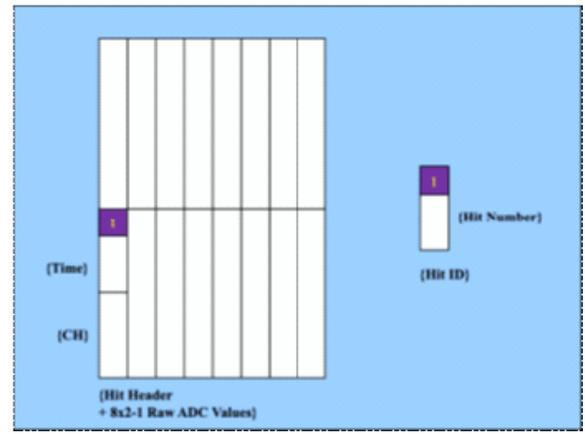

Fig. 3. The hit data structure and the Hit ID RAM content

During the data filling phase, the header of each hit is detected, and an entry is stored in the Hit ID RAM. Each hit occupies one location in the Hit ID RAM. The RAM address is derived from the pair {Time, CH} in the hit header, while the stored content word, as shown in the right of Fig. 3, consists of the Hit Valid flag together with the Hit Number which is simply the running count of the input hit stream.

During the data filling phase, the bits in the Bit Register are set accordingly. Each bit corresponding a hit being filled, and the Priority Encoder always output the highest Hit Number to the Hit Indexing Unit.

The second phase of the clustering operation is the reading phase.

The Hit Indexing Unit controls the readout sequence, and the process starts from the highest Hit Number, as determined by the Bit Register and Priority Encoder.

During the readout sequence, the Hit ID RAM is addressed, and the current hit and its neighboring channels (CH+1 and CH−1) are read out and checked in the Next Hit ID Unit.

If another hit belonging to the same cluster exists, the Hit Indexing Unit continues by reading that next hit.

If no additional hits remain in the cluster, the unit selects the next unread hit with highest Hit Number using the output of the Priority Encoder. The sequence terminates after all hits have been read out.

The reading each hit requires eight clock cycles, and the clock cycle numbers are marked in Fig. 2:

- Clock (0): The Hit ID of the current hit is used to address the Hit Buffer.
- Clock (2): CH and Time bin of the current hit are available at the output port of the Hit Buffer.
- Clock (3, 4): The current hit is addressed at clock cycle (3) and cleared at (4) in the port A of the Hit ID RAM. In the port B, the CH+1 bins are addressed at (3) and the CH-1 bins are addressed at (4).
- Clock (5, 6): The next hit is selected in the Next Hit ID Unit.
- Clock (7): The next Hit ID, if exist, is available to the Hit Indexing Unit.

During this sequence, the bit corresponding to the current hit in the Bit Register is cleared, and as mentioned above, the contents in the Hit ID RAM for current hit are also cleared. This will ensure a hit in the event is readout once and only once.

To ensure the next hit in a cluster can be found, the Hit ID RAM is implemented with four physical RAM blocks,

corresponding to TM[9..8] = 00, 01, 10 and 11, representing four adjacent timing bins. These RAM blocks are addressed with the scheme shown in Fig. 4.

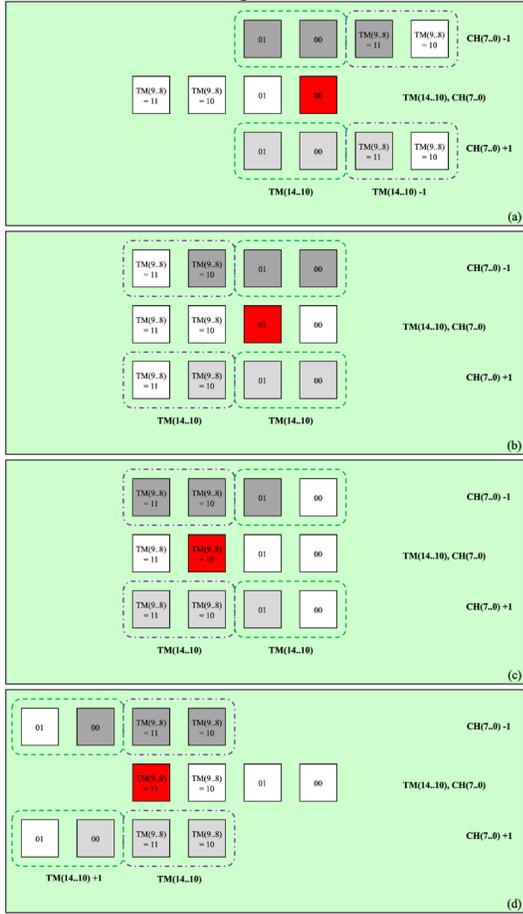

Fig. 4. The Hit ID RAM addressing scheme

In Fig. 4, the middle row representing the port A of the RAM blocks and they are addressed for the current hit. The lower (CH+1) and higher (CH-1) rows represent port B of the RAM blocks, addressed at clock cycle 3 and 4, and readout at clock cycle 5 and 6, respectively. Depending on the value of TM[9..8], the bit field TM[14..10] of port B are assigned with values as shown above. This will ensure in the 2D time-channel map, the bins around the current hit with CH+1 and CH-1 and TM+1, +0 and -1 are searched for the next hit in the cluster.

### III. IMPLEMENTATION IN FPGA

To validate the clustering scheme, described in earlier sections, a clustering engine is implemented in a low-cost FPGA evaluation module and tested. The details of the implementation are described in this section.

The evaluation module is the TERASIC C5G module with a small size, medium speed grade Altera Cyclone V FPGA (5CGXFC5C6F27C7).

The evaluation module is shown in Fig. 5.

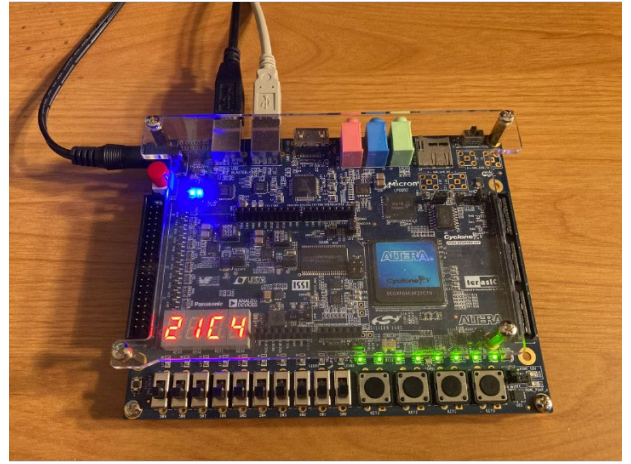

Fig. 5. The FPGA evaluation module used to implement the cluster engine

The clustering engine uses regular FPGA fabric components such as logic elements and block memories and consumes very small amount of resources. The logic paths are carefully designed so that the engine can be successfully compiles with operating frequency of 200 MHz, which is nearly the maximum operating frequency (275MHz for the M10K block RAM) of this FPGA device. The resource usage from the compilation report is listed in TABLE I.

TABLE I
Resource usage of the clustering block and its test system

The clustering engine is implemented and tested in RAM-to-RAM fashion to emulate the real operation conditions. The input data are pre-loaded in the input buffer and then fed through the clustering engine at 200 MHz, the full operation speed. The outputs from the clustering engine are sent into the output buffer and then output through UART port to a notebook computer for analysis.

### IV. TEST RESULTS

The clustering engine are tested with several simulated events with various complexity conditions. One of the events is shown in Fig. 6.

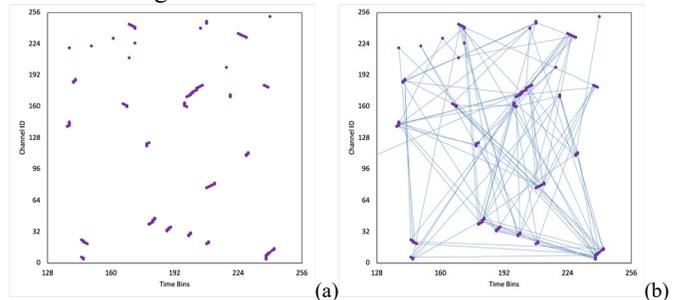

Fig. 6. An event tested in the clustering engine

The plots above represents the 2D time-channel map of an arbitrary TPC event with X-axis being time bin and Y-axis being channel ID. The event shown above has about 110 hits (maximum 128 for the engine in this design) and 28 clusters. The order of the hits in the data stream is random as shown in Fig. 6(b). The straight lines connecting two hits, and the plot shows that the hits belonging to different clusters are blended in the data stream.

After passing through the clustering engine, the hits belonging to the same cluster are reordered and output together as shown in Fig. 7.

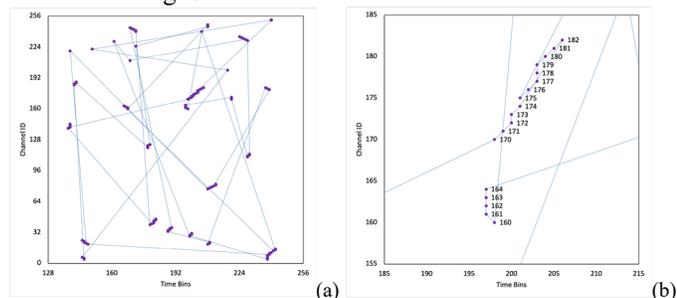

Fig. 7. Output from the FPGA clustering engine

The plot in Fig. 7(a) shows that after clustering, each cluster has only two straight lines connecting other clusters. It indicates that in the output data stream, all hits belonging a cluster are placed together.

The plot shown in Fig. 7(b) shows the details of the data ordering in two clusters. In the higher cluster, the hit with channel ID 175 is output first in the data stream, followed with hits from 176 to 182. Since the hit 182 is the upper end of the cluster, the engine goes back to hit 174 and steps downward to 170 to finish outputting entire cluster. In most cases, the entering point of a cluster is in the middle but occasionally the entering point can be an upper or lower end of the cluster as shown in the cluster from 160 to 164.

To test robustness of the clustering engine with extreme complexity, an event with very long clusters is generated and tested. The data stream of the input and the output of the clustering engine are plotted in Fig. 8(a) and (b), respectively.

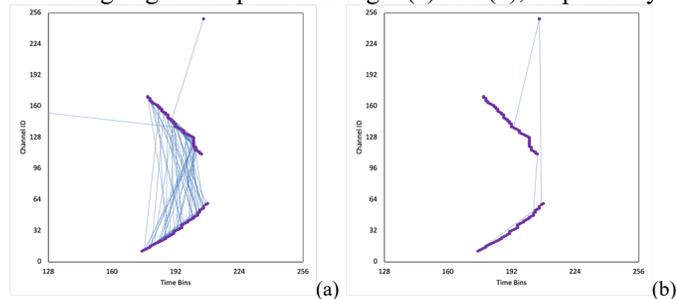

Fig. 8. An event with very long clusters

In the plots shown above, hits from clusters are reorganized correctly. The clustering engine is complexity-agnostic.

As mentioned above, in most cases, the entering hit is in the middle of a cluster, due to that the input data stream is random. This is an imperfectness but is acceptable for many post processing such as finding the center of weight of the cluster. In some applications, however, it is desirable to have clusters ordered in end-to-end fashion. This demand can be realized by simply cascading two cluster engines, processing the data stream twice. After going through the first engine, one end (usually the lower one) of a cluster will be the exiting point, which becomes the entering point in the input to the second engine, and the final output will reorder all clusters from one end to the other (usually from higher to lower end, but there are exceptions depending on the initial order in the input data stream). The output from two cascaded clustering engines in out FPGA test platform is shown in Fig. 9.

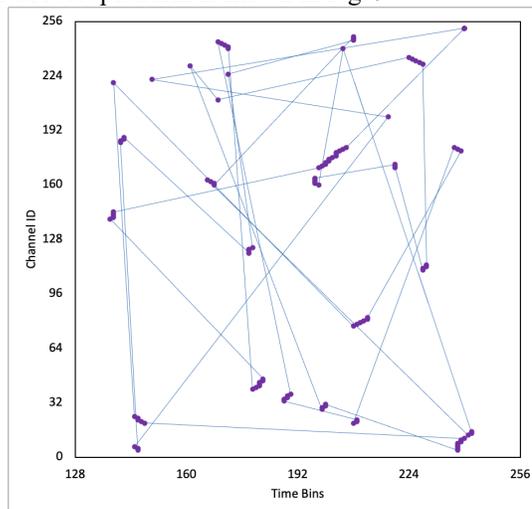

Fig. 9. Output of the cascaded clustering engines

The plot above shows that both the entering and exiting points are at the ends of all clusters, indicating that the hits of a cluster in the data stream are ordered from one end to the other.

## V. DISCUSSIONS

The FPGA devices used in most TPC readout systems for today's high energy physics experiments are much larger than the above one, so it is possible to fit several copies of the clustering engine in most existing hardware. In typical TPC readout systems, ADC sampling rates are relatively low (around 50 MHz for gas filled ones or 2 MHz for liquid Argon ones). The clustering engine described in this document operates 200 MHz system frequency. So it is likely one or two copies of the engines would be sufficient for most systems.
.